\documentstyle[epsfig,amsmath]{mn}
\title[Jet speeds in radio sources]{On the jet speeds of classical double radio sources}
\author[T.\,G.\,Arshakian and
M.\,S.\,Longair]{T.\,G.\,Arshakian$^{1,2}$\thanks{Present address:
Max-Planck-Institute f\"ur Radioastronomie, Auf dem H\"ugel 69, 53121
Bonn, Germany (E-mail: tigar@mpifr-bonn.mpg.de)} and M.\,S.\,Longair$^1$ \\$^1$ Cavendish Astrophysics Group, Cavendish Laboratory, Madingley 
Road, Cambridge, CB3 0HE, United Kingdom \\ $^2$ Byurakan Astrophysical Observatory, Byurakan 378433, Armenia}

\date{ }

\pagerange{\pageref{firstpage}--\pageref{lastpage}}

\pubyear{2003}
\begin{document}

\label{firstpage}

\maketitle

\begin{abstract}
A simple integral relation is obtained for the distribution functions of jet speeds and jet-counterjet flux asymmetries of classical double radio sources on the basis of a simple model in which it is assumed that the jets are intrinsically symmetric and that the asymmetries are attributed to relativistic beaming. Analytic expressions relating the mean jet speed and the mean jet flux asymmetry, and their variances are derived. The results are considered in the light of orientation-based unified schemes, and an analytical equation for the critical angle separating quasars and radio galaxies is derived.  The mean jet speeds of classical double radio sources are estimated using the asymmetry of jet-counterjet flux densities taken from the 3CRR and B2 samples. For FRI radio sources the mean jet speed is $\sim (0.54 \pm 0.03)c$, while for FRII low-redshift radio galaxies and intermediate redshift quasars the values found are $\sim (0.4 \pm 0.06)c$ and $\ga 0.6c$ respectively.
\end{abstract}

\begin{keywords}
active galaxies -- jets -- quasars -- radio continuum: galaxies: theory
\end{keywords}

\section{Introduction}

There is observational evidence that the jets of double FRI and II radio sources are initially relativistic. On parsec scales this idea is supported by the observation of superluminal and relativistic motions, rapid variability and high brightness temperatures. On kiloparsec-scales the correspondence of jet-sidedness with depolarization asymmetry of both FRI (Morganti {\it et al.} 1997) and FRII radio sources (Laing 1988, Garrington {\it et al.} 1988) is interpreted as the brighter jet approaching the observer, thus supporting the idea of relativistic beaming of the radiation of the jet. In all double radio sources in which one-sided parsec and kiloparesc-scale jets are observed, the parsec-scale jet points in the direction of a kiloparsec-scale jet, implying that relativistic speeds persist on kiloparsec scales (Pearson \& Readhead 1988, Wardle and Aarons 1997).

Other observational data are consistent with this picture.  Laing {\it et al.} (1999) analysed the asymmetries in the jet flux densities of 38 FRI radio sources selected from the B2 sample of Parma {\it et al.} (1987) and concluded that jets are intrinsically symmetric and initially relativistic with $\beta\approx 0.9c$. Analysing a large sample of FRII radio galaxies with $z<0.3$ Hardcastle {\it et al.} (1999, hereafter H99) found that relativistic beaming is needed to explain the relationship between core and jet prominences, with speeds between $(0.5-0.6)c$ on kiloparsec scales. Wardle \& Aaron (1997, hereafter W97) analysed the jet-counterjet flux ratios of 13 3CR FRII quasars imaged by Bridle {\it et al.} (1994). They showed that only modest intrinsic jet asymmetries are allowed and that the large jet-counterjet brightness ratios can be attributed almost entirely to the Doppler effect. The idea that jets are intrinsically symmetrical is supported by other observational evidence, for example, the detection of two parsec-scale jets in the giant double radio source DA240 with a flux density ratio of unity (Saripalli {\it et al.} 1997) and the existence of the Compact Symmetric Objects which are thought to be formed by symmetric jets.  Readhead {\it et al.} (1996) and Owsianik \& Conway (1998) argued that Compact Symmetric Objects represent a very early phase of classical double radio sources.

These analyses have assumed that the jet speeds associated with relativistic beaming can be adequately described by a single jet speed for a given source.  This seems to be a reasonable approximation in view of the remarkable symmetry of the hot spots observed in the extended lobes of FRII sources, which need to be constantly supplied with energy by such jets over the lifetime of the source.  There is however an issue whether or not the observed emission from the most compact jets is associated with a single jet speed.  The polarisation structures of the compact jets observed in FRI sources by Laing (1993), Hardcastle et al (1996) and Laing and Bridle (2002) suggest that these jets can be modelled by a spine-shear layer structure, in which entrained material in the shear layer would result in a dispersion in the speeds of the radio emitting material.  These structures are assumed to be fast decelerating jets which do not persist much beyond their observed locations in the radio galaxies.   These effects become important beyond what they term the 'flaring point' of the jets.  Most of the emission from the compact jet originates, however, from the jet spine which is assumed to have a constant speed.   Thus, even in the case of the FRI jets, it would be an adequate assumption that the bulk of the emission is associated with the jet spine, certainly at distances less than the 'flaring point'.   It will be assumed that jet speeds within a given FRII source can be adequately modelled by a single jet speed in the analysis which follows.

To date, estimates of the jet speeds have been found from Monte-Carlo simulations in which the jet properties are determined by relativistic beaming and consistency sought between the data and the model.  In this paper, the reconstruction of the distribution function of jet speeds is found using an inverse-problem approach, in which the mean jet speed and its standard deviation are found directly from the observed jet--counterjet brightness distribution of FRI (Laing {\it et al.} 1999, hereafter L99) and FRII radio sources (Bridle {\it et al.} 1994, Hardcastle {\it et al.} 1998, hereafter B94 and H98).  In Section 2, the inverse problem is formulated and the distribution function of jet speeds and analytical equations for estimating the mean jet speed and its variance are derived.  These results are examined in the context of orientation-based
unified schemes in Section 3, and a new method of estimating the critical angle which separates radio galaxies and radio quasars is derived. In Section 4 and 5 the reduced equations are applied to the data available for FRI and FRII radio sources, and the discussion and conclusions are presented in Section 6. In our computations, Hubble's constant is taken to be $H_0 = 50$ km ${\rm s}^{-1}$ ${\rm Mpc}^{-1}$ and the world model $\Omega_0 = 1, \Omega_{\Lambda} = 0$.

\section{The inverse problem approach}

\subsection{The model}

Although superluminal motions have been interpreted as ``pattern speeds'' rather than the bulk motions of the jets themselves, we adopt the latter picture in the present analysis as the most natural interpretation of the observed asymmetries of the jets since it is entirely consistent with other constraints on viable models for the physics of FRII sources.  The specific assumptions made in the model are as follows.  First, it is assumed that, intrinsically, the jets are symmetric and that the asymmetric observed brightness distribution is entirely attributable to relativistic beaming.  Second, it is assumed that jets are oriented randomly with respect to the line of sight.  The primary selection criterion for the radio sources is that they are chosen from complete flux-density limited samples of extended radio sources at a low radio frequency of 178 MHz.  Since the flux densities of the sources are wholly determined at this frequency by the total flux density of the unbeamed extended emission with no contribution from Doppler-boosted emission, there is no bias in the selection of the radio jets.  Third, we can adopt the procedures we have already discussed to understand the asymmetries of distributions of hot-spots to analyse the statistical properties of the flux density ratios of the radio jets (Arshakian and Longair 2000).    

\subsection{The distribution function of jet speeds}

Assuming that the jets are intrinsically symmetric and that the observed jet flux asymmetries are entirely due to Doppler beaming, the ratio of the jet and counterjet flux densities, $J=S_{\rm j}/S_{\rm cj}$, can be written 
\begin{equation}
 J = \left(\frac{1+\beta_{\rm j}\,\cos\theta}{1-\beta_{\rm j}\,\cos\theta}\right)^{\delta},
\end{equation}
where is $\beta_{\rm j}c$ is the speed of the jet, which is inclined at an angle $\theta \in [0,\pi/2]$ to the line of sight to the observer, and $\delta = m + \alpha$; the constant $m=2$ for a continuous jet (Scheuer \& Readhead 1979).  $\alpha$ is the spectral index, defined by $S_{\nu} \propto \nu^{-\alpha}$ and taken to be 0.6.

Let us invert equation (1) to express $\omega = \beta_{\rm j}\cos\theta$ in terms of $J$,
\begin{equation}
\omega \equiv \frac{J^{1/\delta}-1}{J^{1/\delta}+1} = \beta_{\rm j} \cos\theta,
\end{equation}
Suppose the probability distribution of jet flux asymmetries, as parameterised by $\omega$, is given by the distribution function $g(\omega)$ and is known from observation.  Our task is to determine the distribution function of jet speeds $G(\beta_{\rm j})$ given $g(\omega)$.

According to the model, $\beta_{\rm j}$ and $\cos\theta$ are independent random variables since there is no bias in the selection of the sources and we assume that the jet can be characterised by a single jet speed. Therefore, the cumulative distribution function  $D(\omega) = \int^{\omega}_1 g(x)\,{\rm d}x$ for $\omega$ is,
\begin{equation}
D(\omega)=\int\limits_{\beta_{\rm j}\cos\theta}\int\limits_{\le{\omega}}G(\beta_{\rm j})K(\theta)\,{\rm d}\beta_{\rm j}\,{\rm d}\theta.
\end{equation}
Assuming the radio axes are distributed isotropically over the sky, $K(\theta)=\sin\theta$. Opening the limits of integration,
\begin{equation}
g(\omega)=\int_{\omega}^{1}\frac{G(\beta_{\rm j})}{\beta_{\rm j}}\,{\rm d}\beta_{\rm j},
\end{equation}
and differentiating with respect to $\omega$,
\begin{equation}
G(\omega) = -\,\omega \,g'(\omega).
\end{equation}
This procedure for relating $G(\omega)$ and the derivative of $g(\omega)$ is similar to the reduction carried out by Arshakian and Longair (2000).

\subsection{The mean jet speed and variance}
Multiplying equation (4) by $\omega^k\,{\rm d}\omega$ and integrating from zero to 1, we obtain the $k^{\rm th}$ moment of the distribution $g(\omega)$,
\begin{eqnarray}
\nu_k = \frac{\nu_{0k}}{k+1},
\end{eqnarray}
where $\nu_{0k}$ is the $k^{\rm th}$ moment of the distribution function of true velocities,
\begin{equation}
\nu_{0k}=\int_0^{1}\beta_{\rm j}^kG(\beta_{\rm j})\,{\rm d}\beta_{\rm j}.
\end{equation}
The first and second moments of the distribution function for the jet speeds are equal to the mean speed and the mean square speed of the jet, $\nu_{01} = \overline{v_0}$ and  $\nu_{02}=\overline{\beta_{\rm j}^2}$, while the same moments for the function $g(\omega)$ are  $\nu_1=\overline{\omega}$ and $\nu_2=\overline{\omega^2}$.  Hence, from equation (6), the mean jet speed is given by
\begin{equation}
\overline{\beta}_{\rm j}=2\overline{\omega},
\end{equation}
while the mean squared jet speed is
$\overline{\beta_{\rm j}^2}=3\overline{\omega^2}$.  The dispersion of jet speeds is given by,
\begin{equation}
\sigma_{\beta_{\rm j}}^2=\overline{\beta_{\rm j}^2} - \overline{\beta}_{\rm j}^2 = 3\,\overline{\omega^2} - 4 \,\overline{\omega}^2.
\end{equation}

\section{Unified schemes}

Scheuer (1987) suggested that powerful radio galaxies and radio quasars are drawn from the same distribution of sources, but are viewed at different angles, the quasars having smaller angles to the line of sight than the radio galaxies. There is considerable evidence supporting this idea from analyses of the average sizes, structural asymmetries, misalignment angles and the jet-sidedness of radio quasars and radio galaxies (Barthel 1989, Scheuer 1995, Best {\it et al.} 1995, Arshakian \& Longair 2000). Let us derive the distribution function of jet speeds and the relations between the moments of the distribution functions of jet speeds and $g(\omega)$ within the context of this orientation-based unified scheme.

\subsection{General case}
Suppose the radio axes of certain FRII sources are observed in the range of angles $\theta\,{\in(\theta_1,\theta_2)}$ and that the distribution of angles is random in this interval.  Then, $K(\theta)\,d\theta=C\sin\theta\,d\theta$, where $C=1/(\cos\theta_1-\cos\theta_2)$. Then equation (4) becomes
\begin{equation}
\displaystyle{
g(\omega)=C
\left\{
\begin{array}{cc}
\displaystyle \int\limits_{\omega/\cos\theta_1}^{\omega/\cos\theta_2}\displaystyle\frac{G(\beta_{\rm j})}{\beta_{\rm j}}\,{\rm d}\beta_{\rm j},& \mbox{for } \omega<{\cos\theta_2}; \\
\displaystyle \int\limits_{\omega/\cos\theta_1}^{1}\displaystyle\frac{G(\beta_{\rm j})}{\beta_{\rm j}}\,{\rm d}\beta_{\rm j}, & \mbox{for } \omega\ge{\cos\theta_2},
\end{array}
\right.
}
\end{equation}
under the corresponding conditions that $\beta_{\rm j}\in{[0,1]}$, $\cos\theta\in{[\cos\theta_1,\cos\theta_2]}$ and $\omega\in{[0,\cos\theta_1]}$.

This system of equations was obtained independently by Baryshev and Teerikorpi (1995) in their study of the distribution function of hotspot velocities for FRII radio sources.

\subsection{The mean jet speed and variance}

In the orientation-based unification scheme, the radio galaxies and radio quasars are observed at different angles to the line of sight and are separated by a critical angle $\theta_{\rm c} \sim 45^{\circ}$ (Barthel 1989). For quasars $\theta_1=0$ and $\theta_2=\theta_{\rm c}$, and so equations (10) become, 
\begin{equation}
\displaystyle{
  g(\omega)=C_{\rm Q}
    \left\{\begin{array}{cc}
      \displaystyle\int\limits_{\omega}^{\omega/\cos\theta_{\rm c}}\displaystyle\frac{G(\beta_{\rm j})}{\beta_{\rm j}}\,{\rm d}\beta_{\rm j},
      & \mbox{for }\omega<{\cos\theta_{\rm c}}; \\     
\displaystyle\int\limits_{\omega}^{1}\displaystyle\frac{G(\beta_{\rm j})}{\beta_{\rm j}}\,{\rm d}\beta_{\rm j},& \mbox{for } \omega\ge{\cos\theta_{\rm c}},
\end{array}
\right. 
}
\end{equation}
where $C_{\rm Q} = 1/(1-\cos\theta_{\rm c})$. For the radio galaxies $\theta_1=\theta_{\rm c}$, $\theta_2=\pi/2$ and so,
\begin{equation}
g(\omega)=C_{\rm G}\int\limits_{\omega/\cos\theta_{\rm c}}^{1}\frac{G(\beta_{\rm j})}{\beta_{\rm j}}{\rm d}\beta_{\rm j}, \mbox{ for } \omega\le {\cos\theta_{\rm c}},
\end{equation}
where $C_{\rm G} = 1/\cos\theta_{\rm c}$. For quasars, an analytical solution for the moments can be obtained from the first integral in equations (11),
\begin{equation}
\nu_{0n}=\frac{(n+1)(1-\cos\theta_{\rm c})}{1 - \cos^{(n+1)}\theta_{\rm c}}\,\nu_n.
\end{equation}
Hence the mean speed, mean squared speed and the dispersion are,
\begin{equation}
\nu_{01}\equiv(\overline{\beta}_{\rm j})_{\rm Q}=\frac{2\,(1-\cos\theta_{\rm c})}{1-\cos^2\theta_{\rm c}}\,(\overline{\omega})_{\rm Q},
\end{equation}
\begin{equation}
\nu_{02}\equiv(\overline{\beta_{\rm j}^2})_{\rm Q}=\frac{3\,(1-\cos\theta_{\rm c})}{1-\cos^3\theta_{\rm c}}\,(\overline{\omega^2})_{\rm Q},
\end{equation}
and,
\begin{equation}
(\sigma^2_{\beta_{\rm j}})_{\rm Q}=(\overline{\beta_{\rm j}^2})_{\rm Q} -(\overline{\beta}_{\rm j}^2)_{\rm Q},
\end{equation}
respectively. From equation (12), the corresponding moments for radio galaxies are
\begin{equation}
\nu_{0n}=\frac{n+1}{\cos^n\theta_{\rm c}}\,\nu_n,
\end{equation}
and so,
\begin{equation}
(\overline{\beta}_{\rm j})_{\rm G}=\frac2{\cos\theta_{\rm c}}\,(\overline{\omega})_{\rm G},
\end{equation}
\begin{equation}
(\overline{\beta_{\rm j}^2})_{\rm G}=\frac3{\cos^2\theta_{\rm c}}\,(\overline{\omega^2})_{\rm G},
\end{equation}
and,
\begin{equation}
(\sigma^2_{\beta_{\rm j}})_{\rm G}=\frac1{\cos^2\theta_{\rm c}}\left[3(\overline{\omega^2})_{\rm G}-4(\overline{\omega}^2)_{\rm G}\right].
\end{equation}

If radio galaxies and radio quasars are intrinsically the same objects, the mean jet speed
of radio galaxies and quasars drawn from the same sample should be equal:
$(\overline{\beta}_{\rm j})_{\rm Q+G}=(\overline{\beta}_{\rm j})_{\rm G}$. From equations (8) and (18),
\begin{equation}
  \theta_{\rm c}=\arccos\left[\frac{(\overline{\omega})_{\rm G}}{(\overline{\omega})_{\rm Q+G}}\right].
\end{equation}

\subsection{Test of analytical equations}

In order to test the formalism involving equations (8,9), (14,16), (18,20) and (21), we have modelled the distribution function $g(\omega)$ of jet flux asymmetries from equation (2) by generating 5,000 random values of $\beta_{\rm j}$ and $\theta$ under the assumptions that (i) the distribution function of jet speeds is Gaussian with mean $\overline{\beta}_{\rm m}$ and standard deviation $\sigma_{\beta_{\rm m}}$ and (ii) $\cos\theta$ is uniformly distributed 
in the ranges $[0, 1]$ for equations (8,9), in the range $[0,\cos\theta_{\rm c}]$ for equations (14,16) and in the range $[\cos\theta_{\rm c}, \pi/2]$ for equations (18,20). For various combinations of the free parameters of a model, $\overline{\beta}_{\rm m}$, $\sigma_{\beta_{\rm m}}$ and $\theta_{\rm c}$, the distribution function of $\omega$ has been 
modelled for (i) a joint sample of radio galaxies and quasars, (ii) radio galaxies and (iii) quasars. The mean and mean square values of $g(\omega)$ for these samples are used to calculate the mean jet speed, standard deviation and critical angle using the above analytic equations. We find very good agreement between the different combinations of free parameters and the calculated mean jet speed, standard deviation and critical angle.

\section{Jet speeds in FRII sources}

In order to apply the results of Sect.\,3, high-resolution,
high-sensitivity maps of complete samples of sources are needed.  The
samples of sources used to study the FRII radio sources are listed in
Table 1, which provides statistical data on the total numbers of
sources and the numbers of different types of source within the
samples.  The two samples consist of a sample of high-redshift 3CR
FRII quasars studied by B94 and a sample of low-redshift radio galaxies with $z<0.3$ studied by H98. These studies adopted the same definition of what is meant by a jet and the same method of determining the jet flux densities which are used in the following analysis.

The problem in analysing these data is that the detection of jets is
limited by the sensitivity and resolution of the observations.  In
only a few cases are measured values of $J$ available.  The following
criteria have been adopted. We consider only sources with definite (D),
definite/definite (DD), definite/probable (DP) and probable/probable
(PP) jets defined for the whole jet length, following the terminology
of H98. The sources for which there are only upper limits to the straight jet and counterjet fluxes are classified as `indeterminate' and those in which the straight jet flux density of the brighter jet is less than the straight counterjet flux density were `excluded'. We take the brighter jet to be that approaching the observer and use flux densities from the \emph{straight part of the jet} for estimating the $J$ and hence $\omega$.  The statistics of the jets in the sample are given in Table 1. 

\begin{table*}
\caption{The properties of the radio sources in the B94 and H98 samples} 
\begin{center} 
\begin{tabular} {lcccc} 
\hline
Properties &3CR quasars (B94)& \multicolumn{3}{c}{Low Redshift Radio Galaxies with $z < 0.3$ (H98)} \\
\hline
Total number of sources in sample&13&\multicolumn{3}{c}{44}\\
\hline
&   &  \multicolumn{2}{c}{High excitation radio galaxies}&{Low excitation radio galaxies}\\ 
\hline
Number of sources in sub-sample&13&\multicolumn{2}{c}{32}&{12}\\
Number of sources with detectable jets&13&\multicolumn{2}{c}{22}&{8} \\
Sources for which $J$ can be estimated&12&\multicolumn{2}{c}{17}&{7} \\
Sources for which $J$ is indeterminate&{-}&\multicolumn{2}{c}{15}&{-} \\
Excluded sources& 1    & \multicolumn{2}{c}{5}&{1} \\
\hline
Sources for which $J$ can be estimated  &  & Broad-line & Narrow-line&\\
&&5&12&\\   
\hline
\end{tabular}
\end{center}
\end{table*}

\subsection{FRII radio galaxies}

We analyse the jet velocities of the 32 high-excitation and 12 low-excitation radio galaxies separately, because it has been shown that the FRII low-excitation radio galaxies form a separate group of radio galaxies which are not unified with high-excitation sources (Laing {\it et al.} 1994, Arshakian \& Longair 2000).

H98 state that their low-redshift sample of 44 radio galaxies is free from orientation bias, as the 3CR sample was selected on the basis of a low-frequency (178MHz) survey in which the jets contribute a negligible fraction of the total flux density.   The low redshift limit ($z < 0.3$) means that there are no quasars in the complete sample and so the effects of orientation are associated with a sample of radio galaxies alone.  Of the 32 high-excitation sources, estimates of $J$ could be made for 17 sources -- fluxes were measured from both jets for 5 sources, and only these have accurate values of $\omega$, while for the remaining 12 only lower limits are available. Throughout the paper we treat the lower limits as representing the true values and so this will lead to an underestimate of the mean velocity.  Most of the 15 indeterminate sources are assumed to be inclined at a relatively small angle to the plane of the sky, and are not observable because their flux densities are not Doppler-boosted.  There is therefore a lack of sources at large angles to the line of sight among the 17 for which $J$ values can be estimated.   If we ignore this bias and assume that the 17 radio galaxies have an isotropic distribution, then an upper limit to the mean jet speed can be estimated from equation (8), since values of $J$ close to one are under-represented.  The mean value of $\omega$ for the 17 sources is $\overline{\omega}_{\rm G} =  0.24$ and the mean jet speed is $\le 0.49c$.

Let us estimate the lower limit to the mean jet speed by considering
only those 16 sources of the 32 with one definite jet or
definite/probable jets defined by the whole jet length. We have argued
that these 16 sources are inclined preferentially towards the line of
sight to the observer and so we assume that they are viewed within a
maximum viewing angle $\theta_{\rm c}$ to the line of sight, while the
remaining 16 radio galaxies have orientations between $\theta_{\rm c}$
and $\pi/2$.  An estimate of the critical angle then allows the mean
jet speed and the standard deviation to be calculated from equations
(14, 16). On the assumption that the sources with definite and definite/probable jets are oriented isotropically between $0^{\circ}$ and $\theta_{\rm c}$, we find,
\begin{equation}
\theta_{\rm c} = \arccos \left[ 1 - \frac{N_{\rm D} + N_{\rm DP}}{N_{\rm all}}
\right] \simeq 60^\circ,
\end{equation}
where $N_{\rm D}$ and $N_{\rm DP}$ are the number of sources with
definite or definite/probable jets respectively.

For the 16 high-excitation radio galaxies with D/DP 
jets, the value of $J$ can be estimated only for 11 sources for which the mean value of $\omega$ is $\overline{\omega}_{\rm G} = 0.29$, and so the mean jet speed on kiloparsec scales is $\overline{\beta}_{\rm G} \simeq 0.4 \pm 0.06$ from equations (14) and (16) if $\theta_{\rm c} = 60^\circ$. The mean jet speed lies in the range $(0.35 - 0.43)c \pm (0.06)c$ if $\theta_{\rm c} \in [50^\circ, 70^\circ]$. From this analysis it may be concluded that the mean jet speed of low-redshift FRII radio galaxies is about $0.4c$ with an upper limit of $0.5c$. This range of speeds is slightly lower than those obtained from the analysis of jet prominences by H99, who found $v \approx (0.5-0.7)c$.  Despite the number of upper limits in their analysis, they believe that these values are not far from the true ones.

A similar analysis can be carried out for the sample of 8 definite and 5 possible jets in low-excitation radio galaxies to estimate mean and upper limits to the mean jet speed. The source 3C15 has an excessively large value of $\omega = 0.67$, which is not typical of two-sided jets. Excluding it from the final sample and using equations (8) and (14), the 7 sources result in a probable mean jet speed, $v \ga 0.26c$, and an upper limit, $v \la 0.41c$, if $\theta_{\rm c} \simeq 55^\circ$ for these D/DP jets, assuming that low-excitation radio galaxies form an independent group of sources with a uniform distribution of orientation of the radio axes on the sky. This result suggests that on the average the jet speeds of low-excitation galaxies are slightly smaller than those of high-excitation galaxies, but since it is based on small number statistics, it is only a provisional conclusion.

Arshakian \& Longair (2000) have shown that the lobes of low-excitation radio galaxies have non-relativistic expansion speeds, which are less than the speeds of narrow-/broad-line radio galaxies at the same redshifts. This result is consistent with the lower jet speeds of low-excitation radio galaxies and/or the tendency for them to belong to denser environments (Hardcastle \& Worral 1999) which may be the reason for lower expansion speeds of radio lobes.

\subsection{FRII radio quasars}

The case of the quasars is more complicated.  Quasars with extended radio structures were selected by B94 for VLA observations at 5 GHz and so their radio axes must lie at some angle greater than $\theta_{\rm min}$ to the line of sight.  This inevitably leads to the selection of quasars oriented at angles to the line of sight closer to the upper limit at which they would still be classified as radio quasars, as compared with an unbiased sample -- the sources lie in the range $\theta \in [\theta_{\rm min}, \theta_{\rm c}]$.  For a critical angle in the range $45^{\circ}$ to $60^{\circ}$, W97 estimate the most probable range of jet speeds to be $(0.6-0.7)c$ for 13 quasars observed by B94 by considering the lower limits to the jet-counterjet brightness ratios to represent their true values.  For the same sample, we calculate from equations (14) and (16) the range of mean jet speeds to be $\sim (0.5-0.56)c$ and $\sigma_{\beta_{\rm Q}} \simeq 0.22$, if $45^{\circ} < \theta_{\rm c} <  60^{\circ}$. This result supports the estimates of W97, confirming the presence of highly relativistic speeds in the jets of FRII quasars.

Among the sources in this sample of quasars, analysis of the jet-counterjet flux asymmetries and the luminosity of FRII quasars shows that, for at least two quasars (3C68.1 and 3C208), the jet flux asymmetry cannot be entirely attributed to Doppler beaming and contains a significant fraction of confusing emission from the straight part of the jets (see 
Appendix). This leads to smaller values of $\omega$, which results in an underestimate of jet speeds. Excluding these quasars from the final sample results in a total of 11 FRII quasars. The lower and upper limits to the mean jet speeds can be estimated assuming that the quasars are inclined preferentially, (i) at small angles $(0-\theta_{\rm c})$ and (ii) at large $(\theta_{\rm min}-90^{\circ})$ angles, using equations (14) and (18) respectively. Taking $\overline{\omega} \simeq 0.47$ for all 11 FRII quasars, we find the upper limit to be $\overline{\beta}_{\rm Q} \ge 1$ for $\theta_{\rm min} > 20^{\circ}$ from equation (18), showing directly that the radio quasars cannot be oriented at large angles to the line of sight -- if the quasars were oriented at large angles to the line of sight, there should be many more quasars with double-sided jets with small values of $\omega$.

Several analyses have shown that $\theta_{\rm c} \simeq 45^{\circ}$ for FRII radio sources in the redshift range from $0.5-1$.  8 of the 11 quasars in our sample lie in the redshift range $0.4-1.1$, and so we can adopt the critical angle ($45^{\circ}$) as an upper limit to their inclination angles. From equations (14) and (16), the mean value of $\omega \sim 0.47$ results in a mean jet speed of between $(0.55c - 0.63) \pm 0.05c$, if $\theta_{\rm c} = (45^{\circ} - 60^{\circ})$, which is in fact a lower limit to the mean jet speed. This result demonstrates that the jet speeds on kiloparsec-scales increase from low-redshift FRII radio galaxies, $v \la 0.5c$, to intermediate-redshift FRII quasars, $v \ga 0.6c$.

The speed $v \sim 0.8c$ is obtained by considering the jet flux asymmetries of 5 low-luminosity quasars, 3C 215, 249.1, 334, 336, 351 which all possess two-sided jets and $P < 10^{28.8} {\rm W\,Hz^{-1}}$. These quasars are probably oriented close to a critical angle, $45^{\circ}$, or else two sided jets would scarcely be detectable.  Then, ${\overline{\beta}}_{\rm Q} = 0.56/\cos45^{\circ} \approx 0.8$.

\section{Jet speeds in FRI radio galaxies}

L99 conducted a detailed investigation of 38 weak FRI radio galaxies selected from the B2 sample of Parma {\it et al.} (1987). They defined a sample free from orientation bias by restricting attention to those sources with total flux densities, excluding the radio core, exceeding the survey limit at 408 MHz. All these sources were known to have well-defined straight jets satisfying the criteria of Bridle \& Perley (1984).

Assuming that jets are intrinsically symmetric, they analysed the jet-counterjet brightness ratios at the flaring point of the main jet to estimate the range of jet velocities. The lower limits of the brightness ratio for 10/38 FRI sources were treated as measured values. They analysed the distribution of the logarithm of jet-counterjet flux ratios by fitting it to models using a maximum likelihood approach.   Single-velocity models were found to be unsatisfactory, but a best-fitting velocity was found to be $\beta_{\rm fl} = 0.72$. For their spine/shear-layer model, the common features are a maximum velocity $\sim 0.9c$ and a large dispersion $\ga 0.6c$. They found that the range of velocities at the flaring point is $0.56 \le \beta_{\rm fl} \le 0.8$.

For the same sample, we have calculated the jet properties using equations (8) and (9) assuming that the jets are uniform at the flaring point, that is, our analysis is independent of the jet model. We use the jet flux asymmetry parameter $\omega$ defined by (2) with $\delta = 2.6$. The mean jet speed at the flaring point was found to be $c\overline{\beta}_{\rm fl} = (0.54 \pm 0.03)c$ and the standard deviation of jet speeds ${\sigma_{\beta}}_{\rm fl} = 0.18c$. The higher moments of the distribution determine the skewness and kurtosis of the distribution function of jet speeds, which are ${\rm Skewness} \equiv \overline{\beta_{\rm fl}^3} / (\overline{\beta_{\rm fl}^2})^{3/2} = 4\overline{\omega^{3}}/ (3\overline{\omega^{2}})^{3/2} = 0.64$ and ${\rm Kurtosis} \equiv \overline{\beta_{\rm fl}^4} / (\overline{\beta_{\rm fl}^2})^{2} - 3 = 5\overline{\omega^{4}}/ (3\overline{\omega^{2}})^{2}-3 = -1.7$. The negative kurtosis indicates that the distribution function of jet speeds is flat-topped, while the positive skewness implies that the distribution is asymmetric with the maximum shifted to slightly smaller velocities. The maximum speed, $v \ga 0.72c$, is found for the source 1652+39.  Despite the fact that the mean jet speed is within $3\sigma$ of the high range of speeds $0.56c-0.8c$ obtained by L99, we conclude that most FRI radio sources have jet speeds less than $0.54c$, with a large spread from subrelativistic to high relativistic speeds.

L99 also point out a selection effect leading to a lack of face-on sources because their jets may be too faint to detect or to measure accurately. Inclusion of these sources would decrease the mean jet speed even more, $\overline{\beta}_{\rm fl} < 0.54$.

\section{Conclusions and discussions}

Subject to the simplifying assumptions of the model, which are described in the first paragraph of Section 2.1, the principal results of this analysis are:
\begin{enumerate}
 \item On the basis of an intrinsically-symmetric relativistic jet model, the distribution function of jet speeds is derived, and analytic expressions for the mean jet speed and its standard deviation are obtained. The equations for estimating the same quantities and the critical angle are also derived in the context of orientation-based unification 
schemes.
 \item Applying this analysis to the jet flux asymmetries of double radio sources, we find that:
\begin{enumerate}
 \item[(a)] the mean jet speed of low-redshift FRII radio galaxies lies in the range $\sim (0.4c - 0.5c) \pm 0.05c$ and the standard deviation $\sim 0.15c$; these speeds are in agreement with the range of speeds  obtained from the analysis of jet prominences,
$0.5c-0.7c$ (H99); 
\item[(b)] a lower limit to the mean jet speed of intermediate-redshift FRII quasars is $v \ga 0.6c$, which supports the presence of highly relativistic jets in quasars (B94, W97), showing that the typical jet speed increases with redshift and/or radio luminosity;
\item[(c)] the mean jet speed of FRI radio galaxies at the flaring point is about $0.54c$ with standard deviation $0.18c$; the peak of the distribution function of jet speeds is shifted to
lower speeds.
\end{enumerate}
\end{enumerate}

When the lower limits to $\omega$ are taken as exact values, the analytic methods of the estimation of the mean jet speeds of FRI/FRII sources give values slightly lower than those based on testing (W97, H99) or fitting (L99) models to the data. In the case of low-redshift radio galaxies, this difference can be attributed to the different asymmetry parameters used -- these are the jet/counterjet flux asymmetry and the jet/lobe emission contrast (H99). Despite this, the calculated mean jet speeds are in agreement within the limits of errors. Jet flux asymmetries have been used in analysing the jet speeds
in FRII quasars and FRI radio galaxies. Our estimates of $\overline{\beta}_{\rm j}$ are 
consistent with the values of W97 and only slightly different from the values of L99.

An important issue is the following: what are the real values of the jet flux asymmetries as parameterised by $\omega$? FRI and FRII low-redshift radio galaxies are seen at angles between $0^\circ - 90^\circ$, while most FRII quasars are inclined at angles less than $45^\circ$. Relativistic boosting of the jet
emission is more important for quasars (B94) than for radio galaxies (H98), and hence a difference between the lower limits to the jet flux asymmetry and the real values should on average be higher for quasars than for radio galaxies.  Simulating the effects of deeper
observations, we multiply the lower limits to the jet-counterjet flux ratios of FRI and FRII radio galaxies by 1.5 and estimate the mean speed in the range $\sim (0.6 \pm 0.03)c$ and $\sim (0.5c - 0.6c) \pm 0.05c$.   For FRII quasars, when the limits on the flux ratios
are doubled, a mean speed between $0.7c$ and $0.8c$ is more acceptable.

\subsection*{Acknowledgements}
We are very grateful to Martin Hardcastle for careful reading of the manuscript and valuable comments, to Julia Riley, Robert Laing and Philip Best for useful discussions and comments, and to Margo Aller for her critical reading of drafts of this paper. TGA gratefully acknowledge an ex-quota award from the Royal Society, and the support of the Alexander von Humboldt Foundation for the award of a Humboldt fellowship.  He also thanks his host Anton Zensus for hospitality at the Max-Planck-Institut f\"ur Radioastronomie, Bonn, where the finishing touches to this paper were applied.
\begin{figure*}
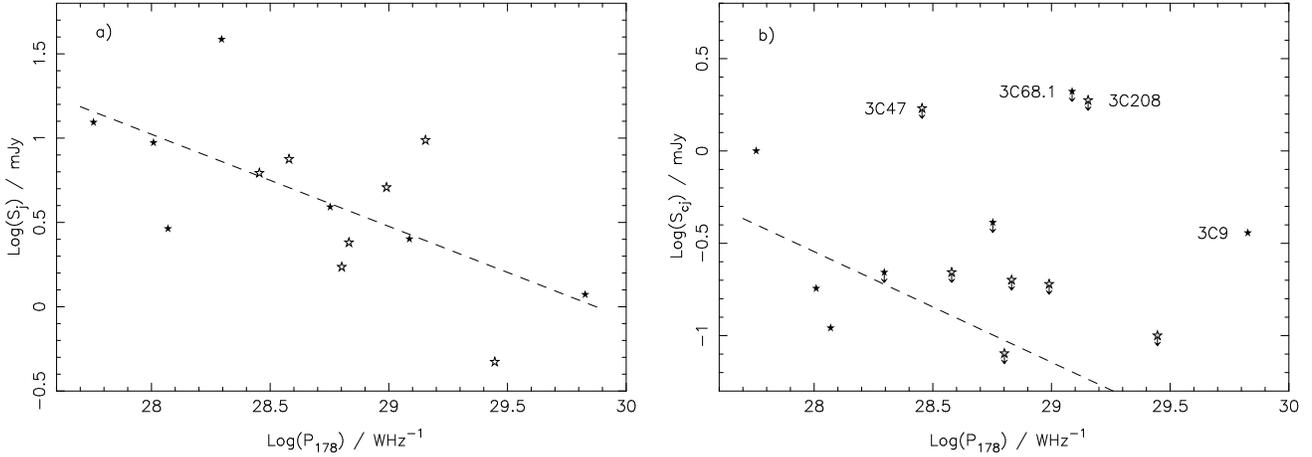

% Malcolm's version:
%\centerline{\epsfig{file = Fig1aRot.eps, width=8cm}
%       \quad\epsfig{file = Fig1bRot.eps, width=8cm}}
% My version:
\centerline{\epsfig{file = Fig1aRot.eps, width=6cm, angle=-90.0}
       \quad\epsfig{file = Fig1bRot.eps, width=6cm, angle=-90.0}}
\caption{(a) Jet and (b) counterjet flux densities of 13 quasars versus the luminosity. Filled asterisks are quasars with two sided jets, while unfilled asterisks are quasars with one sided jets. Upper limits to the straight jet flux are denoted by $\downarrow$. }
\end{figure*}

\subsection*{APPENDIX: EVIDENCE FOR CONFUSING EMISSION IN THE COUNTERJET BRIGHTNESSES OF QUASARS}

There is a negative correlation between the straight jet flux and luminosity, $S_{\rm j} \propto P^{-0.54 \pm 0.18}$ (Fig.\,1a). One should expect that on the average the counterjet flux densities also should correlate negatively with luminosity with some spread depending on the jet speed and jet orientation,
\begin{equation}
S_{\rm cj} = S_{\rm j} \left(\frac{1+\beta_{\rm j}\,\cos\theta}{1-\beta_{\rm j}\,\cos\theta}\right)^{-\delta} \propto P^{-0.54}.
\end{equation}
In fact no correlation is observed as can be seen in Fig.\,1b. A plausible explanation is that some of upper limits to counterjet fluxes densities are overestimated. The expected correlation is shown in Fig.\,1b by the dashed line, which has the slope -0.54 and passes
through the point with coordinates $\log S_{\rm cj} = -0.42$ and $\log P = 28.08$, the mean 
values calculated for four low-luminosity quasars ($P < 10^{28.3}\,{\rm W\,Hz^{-1}}$) with two-sided jets (Fig\,1b). It can be seen that the upper limits to the counterjet brightnesses are overestimated for two quasars, 3C68.1 and 3C208. A plausible interpretation is the presence of confusing emission (B94) in the counterjet lobe, which leads to an overestimation of the counterjet flux density and which in turn seriously underestimates the jet flux asymmetry for these quasars.

%\figure{A1a}(D){4.00cm}{Jet and counterjet flux densities of 13 quasars versus the luminosity, a) and b) respectively. Full asterisks are quasars with two sided jets, while open ones are quasars with one sided jets. The upper limits of straight jet flux asymmetry are denoted by $\downarrow$. }

\label{lastpage}


\begin{thebibliography}{99}
\bibitem{} Arshakian T.\,G., Longair M.\,S., 2000, MNRAS, 311, 846s
\bibitem{} Barthel P.\,D., 1989, ApJ, 336, 606
\bibitem{} Baryshev Yu., Teerikorpi P., 1995, A\&A, 295, 11
\bibitem{} Best P.\,N., Bailer D.\,M., Longair M.\,S., Riley J.\,M., 1995, MNRAS, 275, 1171
\bibitem{} Bridle A.\,H., Perley, R.\,A., 1984, ARA\&A, 22, 319
\bibitem{} Bridle A.\,H., Hough D.\,H., Lonsdale C.\,J., Burns J.\,O., Laing R.\,A., 1994, AJ, 108, 766 (B94)
\bibitem{} Fanaroff B.\,L., Riley J.\,M., MNRAS, 1974, 167, 31P
\bibitem{} Garrington S.\,T., Leahy J.\,P., Conway R.\,G., Laing R.\,A., 1988, Nat, 331, 147
G.\,M., 1997, MNRAS, 288, 859
\bibitem{} Hardcastle M.\,J., Alexander P., Pooley G.\,G., Riley J.\,M., 1998, MNRAS, 296, 445 (H98)
\bibitem{} Hardcastle M.\,J., Alexander P., Pooley G.\,G., Riley J.\,M., 1996, MNRAS, 278,273
\bibitem{} Hardcastle M.\,J., Alexander P., Pooley G.\,G., Riley J.\,M., 1999, MNRAS, 304, 135 (H99)
\bibitem{} Hardcastle M.\,J., Worrall D.\,M., 1999, MNRAS, 909, 969
\bibitem{} Kaiser C.\,R., Alexander P. 1997, MNRAS, 286, 215
\bibitem{} Laing R.\,A., 1988, Nat, 331, 149
\bibitem{} Laing R.\,A. 1993, in Davis R.\,J., Booth R.\,S., eds, Sub-arcsecond Radio Astronomy, Cambridge University Press, p.346 
\bibitem{} Laing R.\,A., Bridle A.\,H., 2002, MNRAS, 336, 328.  
\bibitem{} Laing R.A., Jenkins C.R., Wall J.V., Unger S.W., 1994, in Bicknell G.V., Dopita M.A., Quinn P.J., eds, The first Stromlo Symposium: the Physics of Active Galaxies, ASP
Conference Series vol. 54, p. 201, San Francisco
\bibitem{} Laing R.\,A., Parma P., de Ruiter H.\,R., Fanti R., 1999, MNRAS, 306, 513 (L99)
\bibitem{} Morganti R., Parma P., Capetti A., Fanti R., de Ruiter H.R., 1997, A\&A, 326, 919
J.\,P., Rawlings S., 1995, ApJ, 451, 76
\bibitem{} Owsianik I., Conway J.\,E., 1998, A\&A, 295, 549
\bibitem{} Parma P., Fanti C., Fanti R., Morganti R., de Ruiter H.\,R., 1987, A\&A, 181, 244
\bibitem{} Pearson T.\,J., Readhead, A.\,C.\,S., 1988, AJ, 328, 114
\bibitem{} Readhead, A.\,C.\,S., Taylor G.\,B., Pearson T.\,J., Wilkinson P.\,N., 1996, ApJ, 460, 634
\bibitem{} Saripalli L., Patnaik A.\,R., Porcas R.\,W., Graham D.\,A., 1997, A\&A, 327, 78
\bibitem{} Scheuer P.\,A.\,G., Readhead A.\,C.\,S., 1979, Nat, 277, 182
\bibitem{} Scheuer P.\,A.\,G., 1987, in Zensus J., Pearson T., eds,  Superluminal Radio Sources, Cambridge University Press, Cambridge, p. 104.
\bibitem{} Scheuer P.\,A.\,G., 1995, MNRAS, 277, 331
\bibitem{} Wardle J.\,F.\,C., Aaron S.\,E., 1997, MNRAS, 286, 425 (W97)
\end{thebibliography}
\end{document}